\documentstyle[eqsecnum,twocolumn,prd,aps,epsf]{revtex}

\draft

\begin{document}

\wideabs{

\title{
Gravitational waves from a spinning particle scattered 
by a relativistic star:\\  
Axial mode case}

\author{
Kazuhiro Tominaga$^{1}$
\thanks{Electronic address:tominaga@gravity.phys.waseda.ac.jp},
Motoyuki Saijo$^{2}$
\thanks{Electronic address:saijo@astro.physics.uiuc.edu}, and
Kei-ichi Maeda$^{1,3,4}$
\thanks{Electronic address:maeda@gravity.phys.waseda.ac.jp}
}

\address{
$^{1}$ Department of Physics, Waseda University,
3-4-1 Okubo, Shinjuku, Tokyo 169-8555, Japan
}

\address{
$^{2}$ Department of Physics,
University of Illinois at Urbana-Champaign, \\
1110 West Green Street, Urbana, IL 61801-3080
}

\address{
$^{3}$ Advanced Research Institute for Science and Engineering, \\
Waseda University, Shinjuku, Tokyo 169-8555, Japan
}

\address{
$^{4}$ Max-Planck-Institut f\"{u}r Gravitationphysik,
Albert-Einstein-Institut,\\
Am M\"{u}hlenberg 1, D-14476 Golm, Germany}

\date{August 23, 2000 by K.T.}

\maketitle

\begin{abstract}
We study gravitational waves from a spinning test particle scattered
by a relativistic star using a perturbation method.  The present
analysis is restricted to axial modes. By calculating the energy
spectrum, the waveforms and the total energy and angular momentum
of gravitational waves, we analyze the dependence of 
the emitted gravitational waves on a particle spin. 
For a normal neutron star,  the
energy spectrum has one broad peak whose characteristic frequency
corresponds to the angular velocity at the turning point (a periastron).
Since the turning point is determined by the orbital parameter,
there exists the dependence of the gravitational wave on a particle spin. 
We find that the total energy of $l = 2$
gravitational waves gets  larger as the spin increases in the
anti-parallel direction to the orbital angular momentum.  For an
ultracompact star, in addition to such an orbital contribution, we find
the quasi-normal modes exited by a  scattered particle,  whose
excitation rate
to gravitational waves depends on the
particle spin.  We also discuss the ratio of the total angular
momentum to the total energy of gravitational waves and explain its spin
dependence.
\end{abstract}


}


\section{Introduction}

A neutron star binary  is one of the most promising source for
gravitational waves, which will be observed in the near future  by
gravitational wave interferometers such as Laser Interferometric
Gravitational Wave Observatory (LIGO),  VIRGO, GEO600, TAMA300, and
Laser Interferometer Space Antenna (LISA) \cite{Thorne}.  
Since gravitational waves has a little interaction 
to the matter, we could resolve the phenomena of a strong
gravitational field such as a black hole  formation by the direct
observation of gravitational waves,  which can not be revealed
completely only by electromagnetic waves. From such an analysis, we
could find new information about the interior of a neutron star
whose equation of state is not yet well-known.  These expectations
motivate theoretical efforts to investigate gravitational waves from
a neutron star binary in the inspiral and merging phase.

One of the final step to calculate gravitational waves from a neutron
star binary  is a fully general relativistic simulation with no
approximation. In the past decade, many efforts have been put into the
3D numerical simulation of merging a neutron star binary  in
Newtonian, post-Newtonian, and general relativity.  However, 
Einstein equations are  non-linear wave equations with constraints.
The present simulation has still faced to  many difficulties, however
we hope that we will succeed to simulate some of limited cases in
the near future. In such a situation, we may invoke another way  to
study gravitational waves  from such a binary system by using some
approximation methods focusing to some particular
physical quantities and  extract fundamental and important properties.

One of these approximate approaches of gravitational waves from a
binary system is a perturbation method, in which a companion of the
binary is treated as a test  particle with mass $\mu$ moving around a
relativistic background object with mass $M$ ($\gg \mu$). It has been 
shown that  this approach gives a  good approximation for gravitational
waves from a head-on collision of two black holes \cite{Anninos}, 
and it has been thought to be one of the reliable approximation for treating
gravitational waves from a coalescing black hole binary.

This method can also apply to a neutron star binary, although so far there has
been a strong restriction such that we assume a  spherical
symmetric background spacetime, which can be extended to  a slow rotating
background.  The motion of a test particle excites  nonradial pulsations
of a  central spherically symmetric star in the emission of 
gravitational waves.  The
perturbation equations of gravitational waves outside the spherical star
are given by Regge and Wheeler
\cite{ReggeWheeler} and Zerilli \cite{Zerilli}. Regge-Wheeler-Zerilli
equations are well-known as perturbation equations in the Schwarzschild
background spacetime.  While inside the star, the perturbation equations
both for the metric and stellar fluid were derived by Thorne and
Campolattaro
\cite{ThorneCampolattaro}, Lindblom and Detweiler
\cite{LindblomDeweiler}, and  Chandrasekhar and Ferrari
\cite{ChandrasekharFerrari}.

For a spherically symmetric star, the perturbation equations are
completely separated by its parity, i.e.,  polar (even) and
axial (odd) modes.  The polar mode contains two types of waves;
gravitational waves and material waves  which  interact each other. 
While, the axial mode has only one type, i.e.,  gravitational waves.  
From this character,
quasi-normal modes of fluid oscillations ($f$, $p$ and $g$
modes) are exited by a test particle only in polar modes, while
the quasi-normal modes of gravitational waves ($w$ modes) will appear in
both modes.  As for the axial mode, since there is no fluid oscillation, only  $w$
modes are excited.

In the case of a binary black hole system, it is known that gravitational
waves are dominated  by the quasi-normal modes at the ringing-down
phase. Since the quasi-normal mode contains many information about the
newly formed  black hole or neutron star, the analysis of the
quasi-normal modes is important in the relativistic astrophysics.  As
for a binary neutron star system,  the remnant of merged binary stars
with $M \sim 1.4 M_{\odot}$  may form a black hole after a short
time.  The ringing-down phase of a neutron star binary may be determined
by the quasi-normal modes of a black hole.  However, since the
quasi-normal modes of a star would also be excited at the inspiral phase,
it is still important to focus on them.

For polar modes, Kojima \cite{Kojima} examined gravitational waves
from a particle in a circular orbit around  a spherically symmetric
relativistic star.  From the symmetry of a circular orbit, axial modes
are completely canceled out in this case.  He found the resonant
oscillation of a star  and an enhancement of gravitational waves at the
same frequency of the quasi-normal mode.

For axial modes, Borrelli \cite{Borrelli} calculated the energy
spectrum  when a particle is spiraling onto an ultracompact star
($R \lesssim 3M$).  In the case of the ultracompact star, it is known
that some quasi-normal modes are trapped by a potential barrier of
gravitational waves.  These quasi-normal modes are excited  by a
falling particle.

Several works for a scattered particle have been done.  The present
authors\cite{TominagaSaijoMaeda} calculated the energy spectrum for
axial modes.  In the case of neutron stars ($R \gtrsim 3M$), a single 
peak appears in the energy spectrum, which depends on the trajectory
of a particle.  On the other hand, in the case of ultracompact stars ($R
\lesssim 3M$),  many sharp peaks are also excited in the energy
spectrum.  Some peaks are explained by the quasi-normal modes
trapped by the quasi-bound state in the potential, and the rest of peaks
is due to the resonance of two propagating waves, which are reflected
by the potential barrier or the infinite wall at the center of the star.
These peaks never appear for the black hole case, because there are no
bound state in the potential, and gravitational waves are purely
incoming at the horizon, so the resonance of two reflected waves
never occurs.
Andrade and Price \cite{AndradePrice} also studied axial modes of
gravitational waves by a scattered particle.  They have found that even
for stellar models with $R \gtrsim 3M$, a scattered particle with a
very large orbital angular momentum could weakly excite quasi-normal
modes.
Ferrari, Gualtieri, and Borrelli \cite{FerrariGualtieriBorrelli}
computed energy spectra of both polar and axial modes from a particle
scattered by a polytropic star.  In their results, fluid modes ($f$ and
$p$ modes) are also excited in the energy spectrum.

Recently, there are two papers demonstrated gravitational waves
from a scattered particle in a spherical symmetric star for time
evolution.  Ruoff, Laguna, and Pullin \cite{RLP} computed spectrum
and waveform for polar modes, focusing on the excitation of $w$-modes.
They concluded that the excitation appeared in the spectrum, although
$w$-modes significantly contributes to the energy in the case of the
orbital speed $v > 0.9c$.  Ferrari and Kokkotas \cite{FK} also
computed spectrum and waveform for axial modes.  They also point out
that most of the $w$-modes are indeed, clearly excited in the spectrum
whose feature is the same to our previous results \cite{TominagaSaijoMaeda}.

Neutron stars are usually  rotating. Then it is very natural to
take into account a rotation effect as a next step.  From the
viewpoint of the gravitational wave astronomy, a rotating relativistic
star is a very important object, because there exists $r$ mode which
shows very strong instability for any rotations \cite{r-mode} and this
effect itself becomes one of the promising source of gravitational waves
\cite{OwenLindblomCutlerSchutzVecchioAndersson}.  For example,
hot young neutron stars would be enough rapidly rotating to occur  $r$
mode instability, and would spin down toward a usual observed neutron
star by radiating gravitational waves.  The recent observation
suggests that  low mass X-ray binaries \cite{Backer} would include a
rapidly rotating neutron star.

Although there are several perturbation analysis for a slowly rotating
neutron star, we have no systematic method to solve  for the case of a
rapidly rotating star except for the analysis of the neutral point
\cite{SF}.  Here, we propose an extremely simple model to treat some
effect of a rapid rotation  by adding a spin to a test particle.
Since we can not take into account  the structure of a test particle, we
do not expect the excitation of a test particle itself, and then will not
be  able to discuss the wave mode of a spinning particle itself.  However, this
work might be a milestone for a further study of  a rotation effect on
gravitational waves from  a binary system.

The spinning particle in a curved spacetime was formulated by Papapetrou
\cite{Papapetrou} and Dixon
\cite{Dixon} by taking a point particle limit of a relativistic extended
body.
In the case of a rotating black hole  with a spinning test particle, 
which mimics a collision of two rotating black holes,  
there are several works using a perturbation approach.
\cite{MinoShibataTanaka,TanakaMinoSasakiShibata,SaijoMaedaShibataMino}.

In this paper, we will examine gravitational waves from a spinning
particle scattered by a relativistic star.  We calculate the energy
spectrum, the waveform and the total energy and angular momentum
of the emitted gravitational waves, and investigate the spin effect to
gravitational waves.
This paper is organized as follows.  In Sec. \ref{sec:perturbequation},
we briefly review the equation of motion of a spinning particle and
gravitational waves in a spherically symmetric background
using a perturbation theory.  We show our numerical results in Sec.
\ref{sec:result}.  Section \ref{sec:conclusion} is devoted to discussion.

Throughout this paper, we adopt the units of $G = c = 1$, where $G$ and
$c$ denote the gravitational constant and speed of light respectively,
and the metric signature of $(-,+,+,+)$.

\section{Perturbation equations for axial modes}
\label{sec:perturbequation}

\subsection{Spinning Particle}
\label{subsec:spinparticle}

We briefly summarize the equations of motion of a spinning test particle
and its energy-momentum tensor $T^{\mu \nu}$.  The motion of a spinning
test particle in a curved background is discussed by Papapetrou
\cite{Papapetrou} and Dixon \cite{Dixon}. They presented the  equation of
motions for a spinning particle as
\begin{eqnarray}
\frac{D}{d \tau} p^{\mu}(\tau) &=&
- \frac{1}{2} R^{\mu}{}_{\nu \rho \sigma} (z(\tau))
v^{\nu}(\tau) S^{\rho \sigma}(\tau), 
\label{eq:spin_eom_p} \\
\frac{D}{d \tau} S^{\mu \nu}(\tau) &=&
p^{\mu}(\tau) v^{\nu}(\tau) - p^{\nu}(\tau) v^{\mu}(\tau)
\label{eq:spin_eom_s},
\end{eqnarray}
where $z^{\mu}(\tau)$, $v^{\mu}(\tau) = d z^{\mu} / d \tau$,
$p^{\mu}(\tau)$ and $S^{\mu \nu}(\tau)$ denote the trajectory of a
particle, 4-velocity, 4-momentum, and spin tensor, respectively.  The
mass $\mu$ and the magnitude of a spin $S$, which are
defined by
$p_{\mu} p^{\mu} = - \mu^2$ and $S_{\mu \nu} S^{\mu \nu} = 2 S^2$,
are conserved \cite{Wald}.  Since we also have to specify the center of mass,
which supplies an additional condition, we assume
\begin{eqnarray}
S^{\mu \nu} p_{\nu} = 0,
\end{eqnarray}
with which we find a complete set of equations of motion.  The
normalization of the affine parameter $\tau$ is conveniently chosen as
$u_{\mu} v^{\mu} = -1$, where $u^{\mu} (\equiv p^{\mu} / \mu)$ is the
normalized momentum.

Since the total angular momentum of the particle is conserved, we can
specify the
$z$ axis (and then the equatorial plane) in Schwarzschild spacetime
by the direction of the conserved total angular momentum $\vec{J}$, 
i.e., $\vec{J}=(0,0,J)$ with $J>0$. In this
paper, we restrict  a particle motion onto the equatorial plane
$(\theta = \pi / 2)$ in Schwarzschild background.  
By use of the conserved quantities \cite{Dixon} along the trajectory
$z^{\mu} (\tau) $, we write down  the equation of motion of the
particle with the energy
$\tilde{E} (\equiv E / \mu )$, total angular momentum $\tilde{J} (\equiv
J/ \mu )$  and spin
$\tilde{S} (\equiv S /
\mu )$ \cite{footnote1} as follows \cite{SaijoMaedaShibataMino}:
\begin{eqnarray}
\Sigma_{S} \Lambda_{S} \frac{dt}{d \tau} &=&
\left(1-{2M\over r}\right)^{-1} \left( r^2 \tilde{E} -
\frac{M\tilde{S}}{r} \tilde{J} \right), \\
 \Sigma_{S}^2 \Lambda_{S}^2 \left(\frac{dr}{d \tau} \right)^{2} &=&
R_{S},
\label{eq:spin_eom_r} \\
\Sigma_{S} \Lambda_{S} \frac{d \phi}{d \tau} &=&
\left( 1 + \frac{3 M \tilde{S}^2}{r \Sigma_{S}} \right) \left( \tilde{J}
- \tilde{S} \tilde{E} \right),
\end{eqnarray}
where $\Sigma_{S}$, $\Lambda_{S}$ and $R_{S}$ are
\begin{eqnarray}
\Sigma_{S} &=& r^2 \left( 1 - \frac{M\tilde{S}^2}{r^3} \right), \\
\Lambda_{S} &=& 1 - \frac{3 M \tilde{S}^2 r \left( \tilde{J} - \tilde{S}
\tilde{E}
\right)^2}{\Sigma_{S}^3}, \\
R_{S} &=& \left( r^2 \tilde{E} - \frac{M\tilde{S}}{r} \tilde{J}
\right)^2
\nonumber \\
&&
- r^2 \left(1-{2M\over r}\right)\left[ \frac{\Sigma_{S}^2}{r^2} +
\left( \tilde{J}
- \tilde{S} \tilde{E} \right)^2 \right].
\end{eqnarray}
In what follows, we introduce the ``orbital" angular momentum by
 $\tilde{L} = \tilde{J} -  \tilde{S}$ which is same to Ref. 
\cite{SaijoMaedaShibataMino}, because the orbital angular momentum
plays an important role in the gravitational wave in the case of  a
non-spinning particle \cite{TominagaSaijoMaeda}.  Since we set $J >
0$, the spin direction is classified by the sign of $S$ into the parallel
($S \ge 0$) and anti-parallel  ($S \le 0$)
to the orbital angular momentum.

From Eq. (\ref{eq:spin_eom_r}), we can introduce the effective potential
$\tilde{V}^{\rm (particle)}$ of a spinning particle in the equatorial
plane, which is
\begin{eqnarray}
\tilde{V}^{\rm (particle)}
= \frac{- \beta + \sqrt{\beta^2 - \alpha \gamma}}{\alpha},
\end{eqnarray}
where $\alpha$, $\beta$, $\gamma$ are
\begin{eqnarray}
\alpha &=& r^2 \left[ r^2 - \tilde{S}^2 \left(1 - \frac{2M}{r} \right)
\right], 
\\
\beta &=& r (r - 3M) \tilde{S} \tilde{J}, \\
\gamma &=& \left[ \frac{M^2 \tilde{S}^2}{r^2} - r^2 
\left(1- \frac{2M}{r}\right) \right] \tilde{J}^2
\nonumber \\
&&
- \left(1-{2M\over r}\right) \Sigma_{S}^2.
\end{eqnarray}

\begin{figure}
\begin{center}
\leavevmode
\epsfysize=200pt
\epsfbox{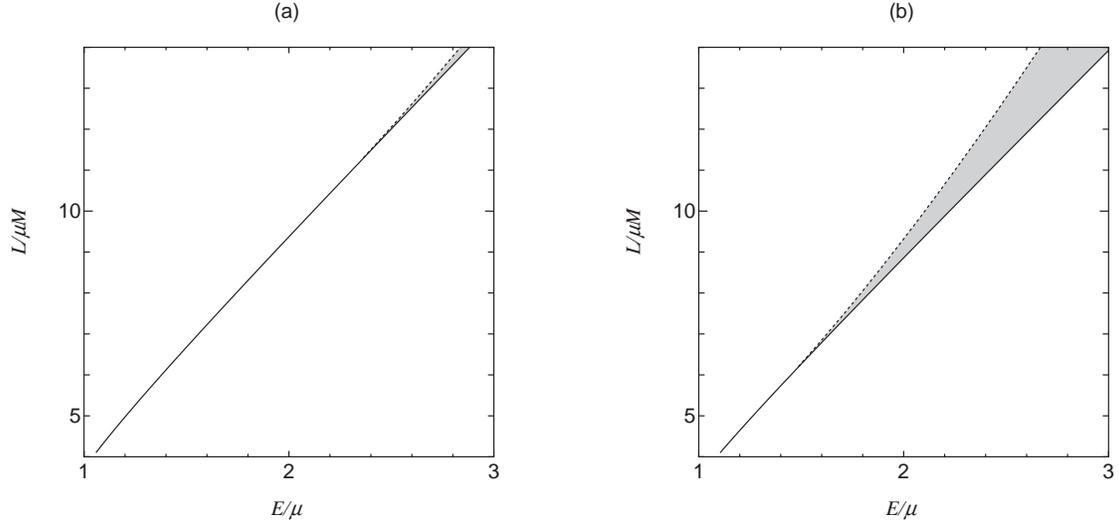}
\end{center}
\caption{
Timelike condition for the 4-velocity and the scattered orbit condition
for a spinning particle are shown in the case of (a) $S = 0.5\mu M$ and
(b) $S = 0.8\mu M$.  The dotted line denotes a null condition, i.e., $v^\mu
v_\mu =0$, and the timelike condition is satisfied in the region above
the dotted line.  The solid line corresponds to the maximum point of
the effective potential $\tilde{V}^{\rm (particle)}$. The shaded region
presents a forbidden parameter for a scattered orbit. 
}
\label{fig:timelike}
\end{figure}

Since we have not normalized the 4-velocity $v^{\mu}$, we may find
some unphysical orbital motion with $v_{\mu} v^{\mu} \ge 0$ for
some specific choice of orbital parameters
\cite{SaijoMaedaShibataMino}. In order not to take such unphysical
parameters, we impose the timelike condition, $v_{\mu} v^{\mu} \le
0$, which is shown in Fig. \ref{fig:timelike}.  The dotted line denotes a
null condition, i.e. $v_\mu v^\mu =0$, and the timelike condition is
satisfied in the region above the dotted line.  The solid line denotes
the maximum point of the effective potential $\tilde{V}^{\rm
(particle)}$. If a particle has smaller energy than that on the solid line
with a given $L$, it will be scattered by a potential barrier, which we
are studying here.  The forbidden parameter range appears as the particle
spin gets larger.

\begin{figure}
\begin{center}
\leavevmode
\epsfysize=200pt
\epsfbox{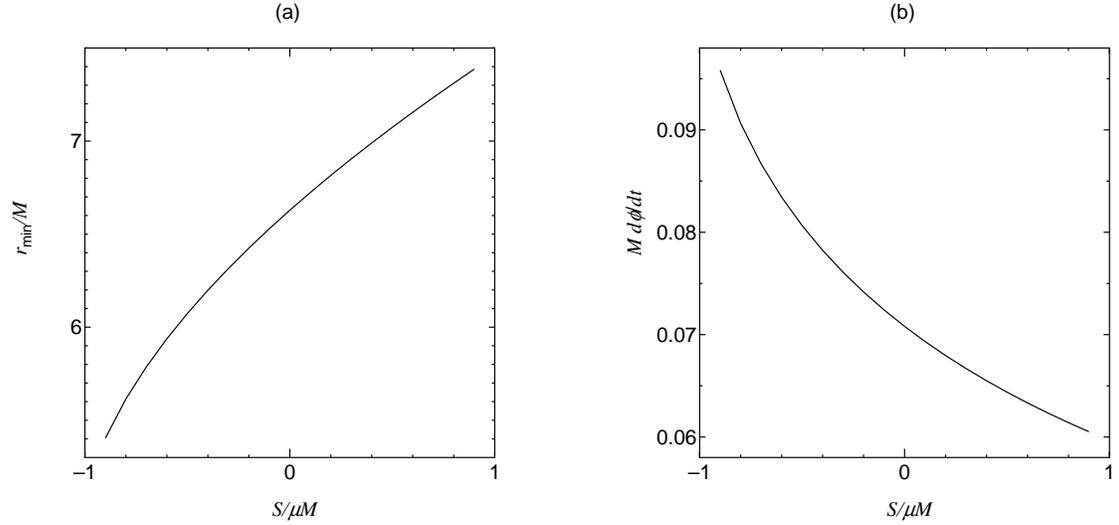}
\end{center}
\caption{
Location (a) and angular velocity (b) at the turning point
of a spinning particle in the case of 
$R = 5.0M$, $E = 1.01 \mu$, and $L = 4.5 \mu M$.
}
\label{fig:turnrm5.0}
\end{figure}

\begin{figure}
\begin{center}
\leavevmode
\epsfysize=200pt
\epsfbox{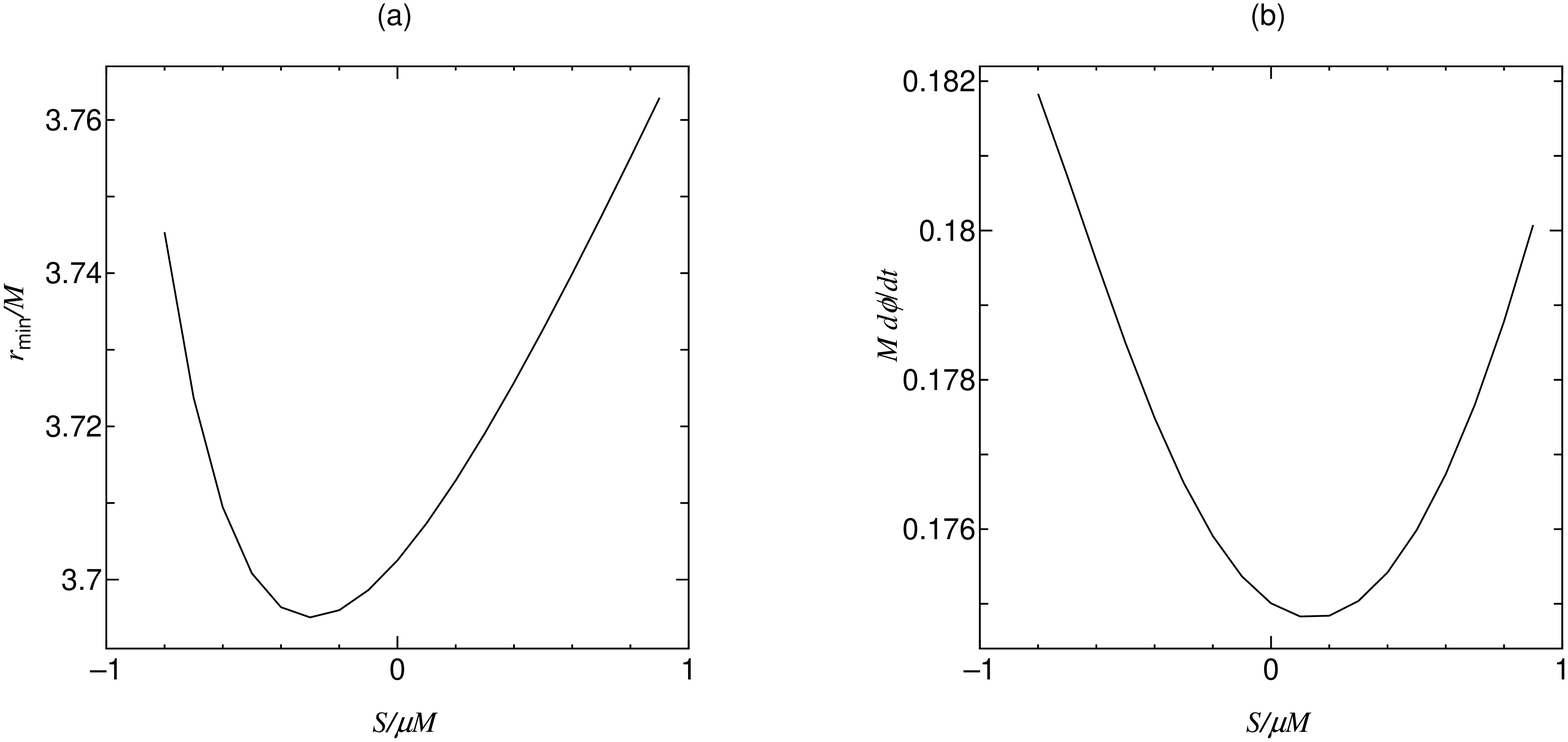}
\end{center}
\caption{
Location (a) and angular velocity (b) at the turning point
of a spinning particle in the case of 
$R = 2.26M$, $E = 2.30 \mu$, and $L = 12.0 \mu M$.
}
\label{fig:turnrm2.26}
\end{figure}

In order to study the emitted gravitational waves from a test particle
 around a relativistic object, it is important to specify the turning
point, i.e. a periastron.  Because the turning point corresponds to the strongest
gravitational field in the particle orbit and its angular velocity gives a
characteristic frequency of the emitted gravitational waves
\cite{TominagaSaijoMaeda}. In Fig. \ref{fig:turnrm5.0},  the
location of the turning point $r_{\rm min}$ and its angular velocity $d
\phi / dt$ are shown in the case of the normal neutron star with $R =
5.0M$. We find that $r_{\rm min}$  increases (and then  $d \phi / dt$
decreases) monotonically as the spin $S$ gets larger in the parallel
direction.  However, if the stellar model is an ultracompact star, e.g.
with $R = 2.26 M$,
and a particle passes through near the ultracompact star,
 the feature at the turning point drastically changes,
and there exists a minimal point for the location and angular velocity
as shown in Fig. \ref{fig:turnrm2.26}. This is just because the particle
passes through a strong
gravitational region, and then nonlinear coupling term between $\vec{L}$ and $\vec{S}$
will appear.

The energy momentum tensor of a spinning particle is given
\cite{MinoShibataTanaka} by
\begin{eqnarray}
T^{\mu \nu} &=& \int d \tau \Biggl[
p^{( \mu}(x, \tau) v^{\nu )} (x, \tau)
\frac{\delta^{(4)} (x - z(\tau))}{\sqrt{-g}}
\nonumber \\
&&
- \nabla_{\rho} \left\{ S^{\rho (\mu} (x, \tau) v^{\nu )} (x, \tau)
\frac{\delta^{(4)} (x - z(\tau))}{\sqrt{-g}} \right\}
\Biggr],
\label{eq:spinparticle_emtensor}
\end{eqnarray}
where $v^{\mu} (x, \tau)$, $p^{\mu} (x, \tau)$ and $S^{\mu \nu} (x,
\tau)$
are extended from the 4-velocity, 4-momentum and spin tensor by the
parallel displacement bi-vector
$\bar{g}^{\mu}{}_{\alpha} (x, x')$ \cite{DeWittBrehme}
as
\begin{eqnarray}
v^{\mu} (x, \tau) &=& \bar{g}^{\mu}{}_{\alpha} (x, z(\tau)) v^{\alpha}
(\tau), \\
p^{\mu} (x, \tau) &=& \bar{g}^{\mu}{}_{\alpha} (x, z(\tau)) p^{\alpha}
(\tau), \\
S^{\mu \nu} (x, \tau) &=&
\bar{g}^{\mu}{}_{\alpha} (x, z(\tau)) \bar{g}^{\nu}{}_{\beta} (x,
z(\tau))
S^{\alpha \beta} (\tau),
\end{eqnarray}
where the bi-vector $\bar{g}^{\mu}{}_{\alpha} (x, z(\tau))$
satisfies
\begin{eqnarray}
&& \lim_{x \to z} \bar{g}_{\mu}{}^{\alpha}(x, z(\tau))
= \delta_{\mu}{}^{\alpha}, \\
&& \lim_{x \to z} \nabla_{\nu} \bar{g}_{\mu \alpha} (x, z(\tau))
= 0.
\end{eqnarray}

\subsection{Linearized Einstein Equation}

We discuss gravitational waves from a spinning test particle scattered
by a relativistic star. We assume that a star is spherically symmetric.
The spherically symmetric background metric is described as
\begin{equation}
g_{\mu\nu}^{(0)} dx^\mu dx^\nu =
-e^{\nu (r)} dt^2 + e^{\lambda (r)} dr^2
+ r^2 \left( d \theta^2 + \sin^2 \theta d \phi^2 \right).
\end{equation}
We shall introduce a mass function $M(r)$ as
\begin{eqnarray}
e^{- \lambda(r)} = 1 - \frac{2 M(r)}{r}.
\end{eqnarray}
The gravitational mass of a star is given by $M = M(R)$ where $R$ is the surface
radius.  We also assume that the background star is composed of a
perfect fluid
\begin{equation}
T^{\mu \nu} = ( \rho + P ) u^{\mu} u^{\nu} + P g^{\mu \nu},
\end{equation}
where $\rho$  and $P$ are the  density and  the pressure of the fluid,
respectively.

In a spherically symmetric background, we can always decompose
the perturbed metric into an axial mode $h_{\mu\nu}^{\rm (axial)}$ and
a polar mode $h_{\mu\nu}^{\rm (polar)}$. In what follows, we discuss only
an axial mode.  In the previous
paper\cite{TominagaSaijoMaeda}, we presented the formalism to calculate
the gravitational wave from a test particle scattered by a star.
In this paper,  we
consider a spinning test particle and discuss the effect of a spin.
Since the formalism is the same except for the source term, which
depends on the character of the particle, we present
a brief sketch of the formalism. We use the same notation and variables
as those in Ref.
\cite{TominagaSaijoMaeda}.

It is known that, using the Regge-Wheeler gauge, perturbation equations
of axial modes are reduced to a  single wave equation both in the
interior and exterior regions of a star.  The perturbation equation
inside a star is given by
\begin{equation}
\frac{d^2 X_{l\omega}^{\rm (int)}}{d r^{*2}}
+ \left( \omega^2 - V_{l}^{\rm (int)} \right) X_{l\omega}^{\rm (int)} =
0,
\end{equation}
where the ``tortoise'' coordinate $r^{*}$  and the effective potential
$V_{l}^{\rm (int)}$ are defined as
\begin{eqnarray}
r^{*} &=& \int_{0}^{r} e^{- (\nu - \lambda)/2} dr, \\
V_{l}^{\rm (int)} &=& e^{\nu}
\left( \frac{l(l+1)}{r^2} - \frac{6M(r)}{r^3}
- 4 \pi (P - \rho) \right),
\label{eq:gw_potential_in}
\end{eqnarray}
respectively.

The perturbation equation outside a star is derived as
\begin{eqnarray}
\frac{d^2 X_{lm\omega}^{\rm (ext)}}{d r^{*2}}
+ \left( \omega^2 - V_{l}^{\rm (ext)} \right) X_{lm\omega}^{\rm (ext)}
= {\cal S}_{lm\omega}^{\rm (ext)},
\label{eq_ext}
\end{eqnarray}
where the tortoise coordinate $r^{*}$ and the Regge-Wheeler potential
$V_{l}^{\rm (ext)}$ are 
 given by
\begin{eqnarray}
r^{*} &=& r + 2M \ln \left( \frac{r}{2M} - 1 \right),
\\
V_{l}^{\rm (ext)} &=&
\left(1-{2M\over r}\right) 
\left( \frac{l(l+1)}{r^2} - \frac{6M}{r^3} \right),
\label{eq:rwpotential}
\end{eqnarray}
respectively.
The source term 
${\cal S}_{lm\omega}^{\rm (ext)}$ in Eq. (\ref{eq_ext}) is
given  by the energy-momentum tensor of a spinning test particle 
(Eq. (\ref{eq:spinparticle_emtensor})) 
expanded by tensor harmonics, whose explicit description is 
presented  in the Appendix (see also Eq. (2.24) in Ref
\cite{TominagaSaijoMaeda}).

To construct the wave functions, $X_{l\omega}^{\rm (int)}$ and
$X_{lm\omega}^{\rm (ext)}$, we have to impose a set of boundary
conditions.  The wave function $X_{l \omega}^{\rm (int)}$
inside a star should be 
regular at the center of a star, which implies
\begin{eqnarray}
X_{l\omega}^{\rm (int)} &=&
\eta_{l\omega} r^{l+1} \Biggl[ 1 + \frac{1}{2(2l+3)}
\nonumber \\
&&
\times \left\{ 4 \pi (l + 2)
\left( \frac{1}{3} (2l - 1) \rho_{\rm c} - P_{\rm c} \right)
- \omega^2 e^{- \nu_{\rm c}} \right\} r^{2} 
\nonumber \\
&&
+ O(r^{4}) \Biggr],
\end{eqnarray}
where $\eta_{l\omega}$ is an arbitrary constant, and $\rho_{\rm c}$,
$P_{\rm c}$, and $\nu_{\rm c}$ are the central values of the density
$\rho$, pressure $P$, and metric function $\nu$, respectively.  
For the wave function
$X_{lm\omega}^{\rm (ext)}$ outside a star, there are no incoming waves at infinity,
which is guaranteed by the condition
\begin{equation}
X_{lm\omega}^{\rm (ext)} \to A_{lm\omega} e^{i \omega r^{*}}
\qquad (r^{*} \to \infty),
\end{equation}
where $A_{lm\omega}$ is the amplitude of an outgoing wave at infinity.

With those boundary conditions, we impose the matching condition  
at the surface $r^{*}=R^{*}$ as
\begin{eqnarray}
X_{l\omega}^{\rm (int)}(R^{*}) &=& X_{lm\omega}^{\rm (ext)}(R^{*}),
\label{eq:S_condition1} \\
 \frac{dX_{l\omega}^{\rm (int)}}{dr^{*}} \left( R^{*}\right) &=&
 \frac{dX_{lm\omega}^{\rm (ext)}}{dr^{*}}\left( R^{*}\right), 
\label{eq:S_condition2}
\end{eqnarray}
which guarantees that the wave function  is continuous and smooth.
In this way, we construct the wave function $X_{lm\omega}$.

Finally, we introduce important physical quantities of gravitational
waves, i.e.,  the total energy, the total angular momentum, the energy spectrum,
the angular momentum spectrum, and the waveform as
\begin{eqnarray}
E_{\rm GW}^{\rm (axial)} &=& 
\int_{- \infty}^{\infty} d\omega \sum_{l,m}
\left(  \frac{dE_{\rm GW}}{d\omega} \right)_{lm\omega}^{\rm (axial)}
,\\
J_{\rm GW}^{\rm (axial)} &=&
\int_{- \infty}^{\infty} d\omega \sum_{l,m}
\left(  \frac{dJ_{\rm GW}}{d\omega} \right)_{lm\omega}^{\rm (axial)}
,\\
\left( \frac{dE_{\rm GW}}{d \omega} \right)_{lm\omega}^{\rm (axial)}
&=&
\frac{1}{32 \pi} l (l + 1) (l - 1) (l + 2) 
\nonumber \\
&& \times
\left|A_{lm\omega}^{\rm (axial)} \right|^2
,
\\
\left( \frac{dJ_{\rm GW}}{d \omega} \right)_{lm\omega}^{\rm (axial)}
&=&
\frac{m}{\omega} \left( \frac{dE_{\rm GW}}{d \omega} \right)_{lm\omega}^{\rm (axial)}
,
\\
h_{+}^{\rm (axial)} \pm i h_{\times}^{\rm (axial)} &=&
\mp \frac{1}{2 \pi r} \int_{- \infty}^{\infty} \frac{d
\omega}{\omega } 
\sum_{l,m}  A_{lm \omega}^{\rm (axial)} \nonumber \\
&& \times \sqrt{l (l + 1) (l - 1) (l + 2)} e^{-i \omega (t - r^{*})} \nonumber \\
&& \times {}_{\pm 2} Y_{lm}
(\theta, \phi),
\end{eqnarray}
where ${}_{\pm 2} Y_{lm}$ is the spin-weighted spherical harmonics.

\section{Gravitational waves from a scattered spinning particle}
\label{sec:result}

We analyze the emitted gravitational waves from a spinning test 
particle scattering around a relativistic star.  Because the potential of
the linearized Einstein equation takes a maximum value around $r \sim
3M$, our stellar model can be classified into two models:  One model is
a normal neutron star  ($R \gtrsim 3M$) whose potential decreases
monotonically as the coordinate radius increases. The observed
neutron stars are classified into this case.  The other model is an
ultracompact star ($R
\lesssim 3M$). It would be very exotic, but show many interesting
relativistic effects, e.g. $w$ modes trapped by a  quasi-bound state of
the potential (see, for example Figs. 5 and 6 in Ref.
\cite{TominagaSaijoMaeda}).  We only discuss most simple equation of state, i.e., 
a uniform density star ($\rho = {\rm const.}$), because the fluid
oscillation will not couple directly to the axial modes of gravitational
waves and then the information of the stellar matter may
not be important.

\subsection{Normal neutron star}

\begin{figure}
\begin{center}
\leavevmode
\epsfysize=200pt
\epsfbox{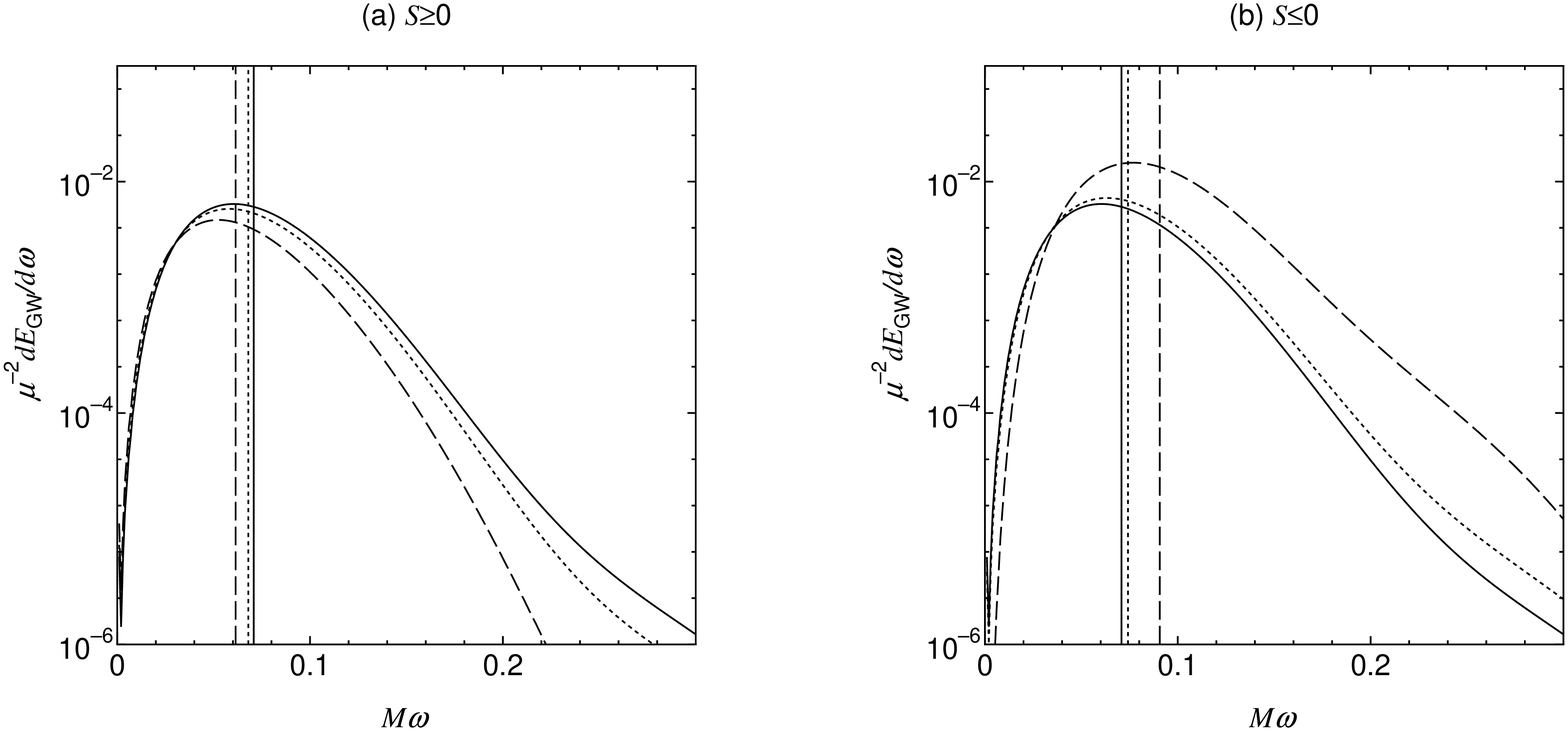}
\end{center}
\caption{
Energy spectrum ($l = 2$ mode) of gravitational waves from a spinning
particle with $E = 1.01 \mu$ and $L = 4.5 \mu M$ scattered by a
uniform density star with $R = 5.0M$. In the parallel case (a), solid,
dotted and dashed lines show the case of $S = 0$, $0.2\mu M$, and
$0.8\mu M$, respectively, while in the anti-parallel case (b), solid, dotted
and dashed lines show $S = 0$, $-0.2\mu M$, and $-0.8\mu M$. 
Vertical lines represent the angular velocity of a particle at the turning
point.
}
\label{fig:specrm5.0}
\end{figure}

Here, we discuss a normal neutron star  with $R \gtrsim 3M$.  In the
previous analysis of a non-spinning particle
\cite{TominagaSaijoMaeda}, we showed that one broad peak appears in
the energy spectrum of gravitational waves, which corresponds to the
angular velocity  of a test particle at the turning point, i.e.,
a periastron.  Since the orbit of a test particle is affected by its spin,
we expect that the existence of a spin will modify the spectrum of
the emitted gravitational waves.  In fact for the case of a spinning particle 
(Fig. \ref{fig:specrm5.0}), we find a broad spectrum quite similar  
to the non-spinning case, 
but we see that the energy spectrum is affected by a particle spin.
The force which acts on the particle is given by 
\begin{eqnarray}
F^{\rm (particle)} &\equiv &
- \mu \frac{\partial \tilde{V}^{\rm (particle)}}{\partial r} 
\nonumber \\
&\sim &
\frac{\mu}{M}  
\Biggl[ 
- \left(\frac{M}{r} \right)^{2} +
\frac{1}{M^{2}}( \tilde{L}^{2} - M^2 ) \left( \frac{M}{r} \right)^{3} 
\nonumber \\
&&
- \frac{3}{2M^{2}} ( \tilde{L}^{2}-4 \tilde{L} \tilde{S} + M^2 )
\left( \frac{M}{r} \right)^{4} \nonumber \\
&& + O \left( \frac{M}{r} \right)^{5}
\Biggr].
\nonumber  
\\
\label{eqn:force}
\end{eqnarray}
As we can see from this expression, the coupling between a
spin
$\vec{S}$ and an orbital angular momentum $\vec{L}$, i.e. $\vec{L} \cdot \vec{S}$
coupling, gives an repulsive force.
As a result,
the turning point leaves away from a stellar surface for a large positive (parallel)
spin and then  the angular velocity of a particle at the turning point
gets smaller (Fig. \ref{fig:turnrm5.0}). In fact,  for a large positive
(parallel) spin, a broad peak in the spectrum shifts to the lower
frequency region, and for a large negative (anti-parallel) spin  
it shifts to the higher frequency region (Fig. \ref{fig:specrm5.0}).   
The change 
of the spectrum strength also corresponds to the shift of the turning
point location.  The amount of gravitational waves increases when the
particle gets a large negative (anti-parallel) spin, because the  gravitational
field becomes strongest at the turning point
and the most amount of gravitational waves radiates at this point.

\begin{figure}
\begin{center}
\leavevmode
\epsfysize=200pt
\epsfbox{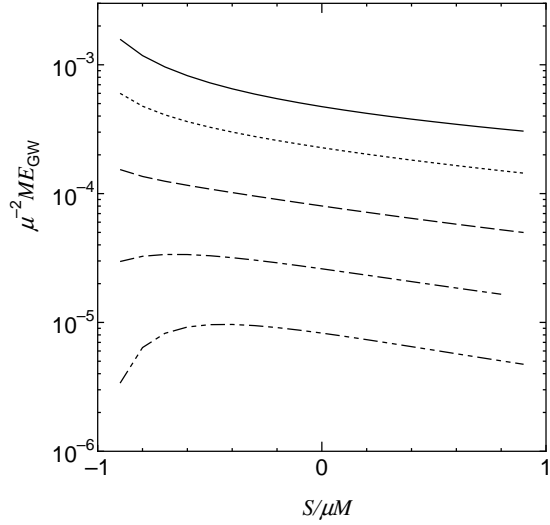}
\end{center}
\caption{
Total energy of gravitational waves for each multipole mode $l$ in the
same situation as Fig. \ref{fig:specrm5.0}. Solid, dotted, dashed,
dash-dotted and dash-two dotted lines show the case of $l = 2, 3, 4, 5,
6$, respectively.
}
\label{fig:totalerm5.0}
\end{figure}

The total energy of gravitational waves is shown in Fig.
\ref{fig:totalerm5.0} for each multipole mode $l$. As we expect, the
$l=2$ mode is dominant and the amount decreases as $l$ gets larger.
As for the spin dependence, we find that for small multipole modes ($2
\le l \le 4$) the total energy monotonically increases as the spin gets
large negative (anti-parallel).  While for
large  multipole modes ($l \ge 5$) such a simple relation disappears,
but the energy reaches some maximum value for some negative value
of $S$. We would expect that there is a complicated coupling in the
higher order of Eq. (\ref{eqn:force}) and its effect might directly
appear to the total energy of gravitational waves.

\begin{figure}
\begin{center}
\leavevmode
\epsfysize=200pt
\epsfbox{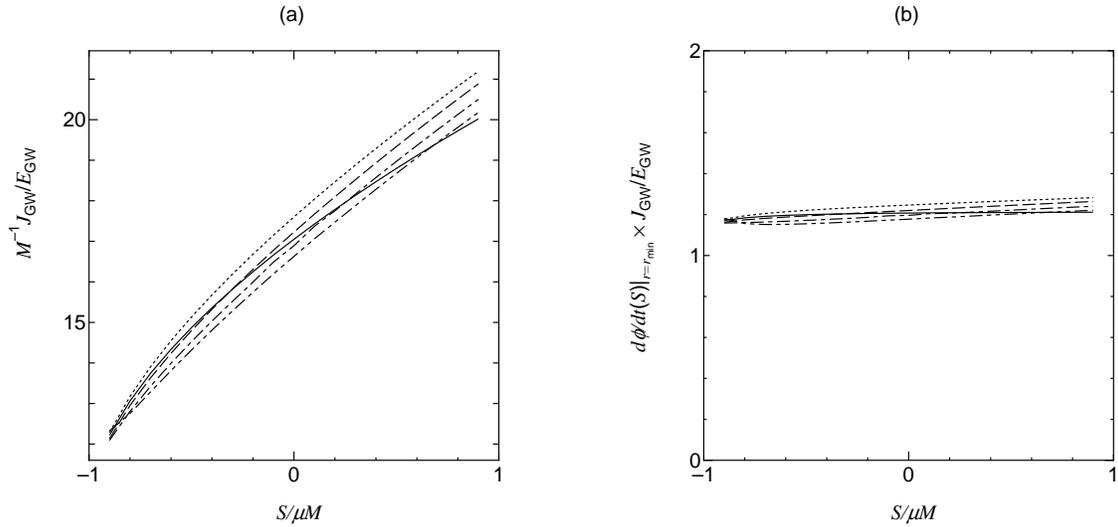}
\end{center}
\caption{
Ratio of the total angular momentum to the total energy radiated as
gravitational waves for each multipole mode $l$ in the same situation
as Fig. \ref{fig:specrm5.0}.  Figure (b) is
the ratio (a) multiplied by 
the angular velocity at the turning point for each spin value, 
in order to estimate Eq. (\ref{eq:J_E_relation}).
Solid, dotted,
dashed, dash-dotted and dash-two dotted lines show the case of $l = 2,
3, 4, 5, 6$, respectively.
}
\label{fig:totaljerm5.0}
\end{figure}

We show the ratio of the total angular momentum to the total
energy, i.e. $J_{\rm GW} / E_{\rm GW}$, in Fig. \ref{fig:totaljerm5.0}
(a).  Remarkably, we  find a simple relationship that $J_{\rm GW} /
E_{\rm GW}$ is proportional to the particle spin, and  depends less on
$l$.  For the case of the scattered particle, the characteristic
frequency of gravitational waves is described by the angular velocity
at the turning point, i.e., 
\begin{equation}
\omega_{\rm ch} = m \left( \frac{d\phi}{dt} \right)_{r=r_{\rm min}}.
\label{eqn:NormalRoughEstimation}
\end{equation}
We roughly estimate the total energy and the total angular
momentum of gravitational waves as
\begin{eqnarray}
(E_{\rm GW})_{l}^{\rm (axial)} &=& 
\int d\omega \sum_{m}
\left( \frac{d E_{\rm {GW}}}{d \omega} \right)^{\rm (axial)}_{lm \omega}
\nonumber\\
&\sim&
\Delta \omega_{\rm ch}
\sum_{m}
\left( \frac{d E_{\rm {GW}}}{d \omega} \right)^{\rm (axial)}_{lm \omega_{\rm ch}}
\label{eqn:Eestimate}
, \\
(J_{\rm GW})_{l}^{\rm (axial)} &=& 
\int d\omega \sum_{m}
 \frac{m}{\omega}
\left( \frac{d E_{\rm {GW}}}{d \omega} \right)^{\rm (axial)}_{lm \omega}
\nonumber\\ 
&\sim& 
\left( \frac{d\phi}{dt} \right)^{-1}_{r=r_{\rm min}}
\Delta \omega_{\rm ch}
\sum_{m}
\left( \frac{d E_{\rm {GW}}}{d \omega} \right)^{\rm (axial)}_{lm \omega_{\rm ch}},
\nonumber \\
\label{eqn:Jestimate}
\end{eqnarray}
where $\Delta \omega_{\rm ch}$ is the width of the broad peak.  Using Eqs.
(\ref{eqn:Eestimate}) and (\ref{eqn:Jestimate}), we find the relation 
\begin{equation}
\left( \frac{J_{\rm GW}}{E_{\rm GW}} \right)_{l}^{\rm (axial)}
\sim 
\left( \frac{d\phi}{dt} \right)^{-1}_{r=r_{\rm min}}.
\label{eq:J_E_relation}
\end{equation}
In fact, our numerical result has a good coincident with this rough
estimation (Fig. \ref{fig:totaljerm5.0} (b)).
We then conclude that  the spin dependence on
$(J_{\rm GW}/E_{\rm GW})_{l}$ comes from the angular velocity
at the turning point.

\begin{figure}
\begin{center}
\leavevmode
\epsfysize=200pt
\epsfbox{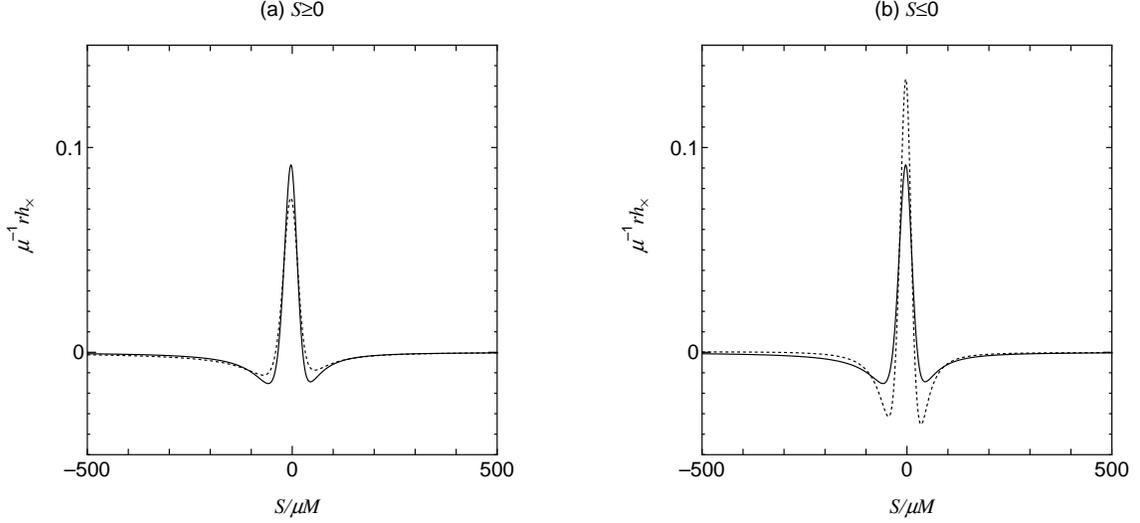}
\end{center}
\caption{
Waveform ($l = 2$ mode) of gravitational waves in the direction of
$\theta = \pi / 2$, $\phi = 0$ in the same situation as Fig.
\ref{fig:specrm5.0}. In the parallel case (a), solid and dotted lines show
the case of $S = 0$ and $0.8\mu M$, respectively, while in the anti-parallel
case (b), solid and dotted lines show $S = 0$ and $-0.8\mu M$.  
}
\label{fig:wfrm5.0}
\end{figure}

As for the waveform, a particle spin does not make a drastic change of
the global feature of it (Fig. \ref{fig:wfrm5.0}).
The difference depending on the particle spin is only the amplitude of the burst
wave. Since the burst wave is generated when the particle comes
to the turning point, the amplitude is more enhanced when the particle can
move in the stronger gravitational field.  This is the reason why 
the amplitude is larger  when the particle has an anti-parallel spin
with the fixed orbital angular momentum.

We conclude for the case of a normal neutron star that the effect of a
spin could be understood just by the shift of the turning point, which
is mainly caused by $\vec{L}\cdot\vec{S}$ coupling.

\subsection{Ultracompact star}

If a star is sufficiently small as an ultracompact star  ($R \lesssim
3M$), the effective potential of the gravitational waves  (Eqs.
(\ref{eq:gw_potential_in}) and (\ref{eq:rwpotential})) has a minimum
and the gravitational waves could be
trapped and enhanced as a quasi-normal mode. This provides an
additional characteristic structure in the spectrum of the emitted
gravitational waves. In the previous works for a non-spinning particle
\cite{TominagaSaijoMaeda,AndradePrice}, we found that a test particle
scattered by an ultracompact star excites many quasi-normal modes, and there appear many sharp peaks
in the energy spectrum, besides one broad peak corresponding to the turning point. 
We also found that such excited quasi-normal
modes will leak later from the potential barrier and the resonance
waves appears periodically  in the waveform. Therefore in the case of
an ultracompact star, there are two aspects in the spectrum: One is a
broad peak which depends on the particle orbit and the other is many sharp
peaks which  depend on the structure of the background star. Here, with
respect to the effect of a spin on the above properties,  we shall
discuss each in order.

\begin{figure}
\begin{center}
\leavevmode
\epsfysize=200pt
\epsfbox{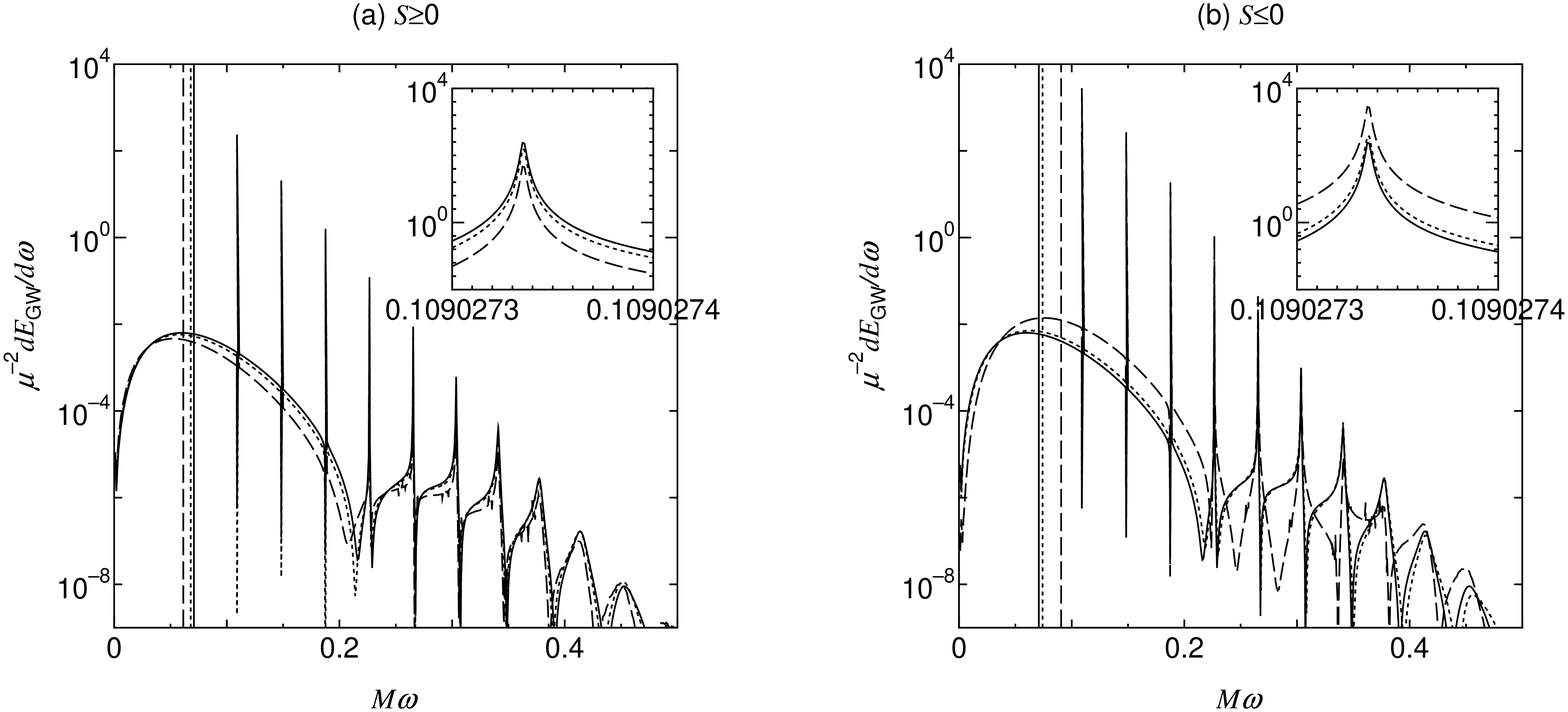}
\end{center}
\caption{
Energy spectrum ($l = 2$ mode) of gravitational waves from a spinning
particle with $E = 1.01 \mu$ and $L = 4.5 \mu M$ scattered by a
uniform density star with $R = 2.26M$. In the parallel case (a), solid,
dotted and dashed lines show the case of the spin $S = 0$, $0.2\mu M$,
and $0.8\mu M$, respectively, while in the anti-parallel case (b), solid,
dotted and dashed lines $S = 0$, $-0.2\mu M$, and $-0.8\mu M$.
Vertical lines represent the angular velocity of a particle at the turning
point.
The small figure in (a) and (b) is the enlargement of 1st quasi-normal mode, 
in order to show the difference of its strength.
}
\label{fig:specfarrm2.26}
\end{figure}

If the  turning point is far from a stellar surface, we find the similar
spectrum to the case of a normal neutron star except for many sharp
peaks. The energy spectrum is shown in Fig. \ref{fig:specfarrm2.26},
in which a particle with $E = 1.01 \mu$ and $L = 4.5 \mu M$ is
scattered at $r_{\rm min} \simeq 5.61M$ ($S = -0.8 \mu M$) and $r_{\rm
min} \simeq 7.31M$ ($S = 0.8 \mu M$) by an ultracompact star with $R =
2.26M$.  As the case of the normal neutron star, the broad peak
shifts to the lower frequency region for a large positive (parallel) spin, while
to the higher frequency region for a large negative (anti-parallel) spin.

We also find many sharp peaks, which correspond to trapped
quasi-normal modes induced by a scattered particle. Since the
quasi-normal modes depend only on a star, not on the property of a
test particle, the frequency of those peaks completely corresponds for any values of a spin. 
The difference appears only in the strength of these peaks (see the small 
figure in Fig. \ref{fig:specfarrm2.26},
which is enlarged on the frequency region of 1st quasi-normal mode). 
In the high frequency region, however, we
also find some peaks in the spectrum, which show a little dependence on
the particle spin. This is because those peaks do not really correspond
to the quasi-normal mode, but appear as a result of an interference
effect between incoming and outgoing gravitational
waves,  which depends on a spin,  reflected by the potential
barrier around $r \sim 3M$ or the infinite wall at the center of a star (see Ref.
\cite{TominagaSaijoMaeda}). However, since the contribution of such peaks to
the emitted gravitational waves is very small,  we may not be able to
distinguish a spin effect in the gravitational waves.

\begin{figure}
\begin{center}
\leavevmode
\epsfysize=200pt
\epsfbox{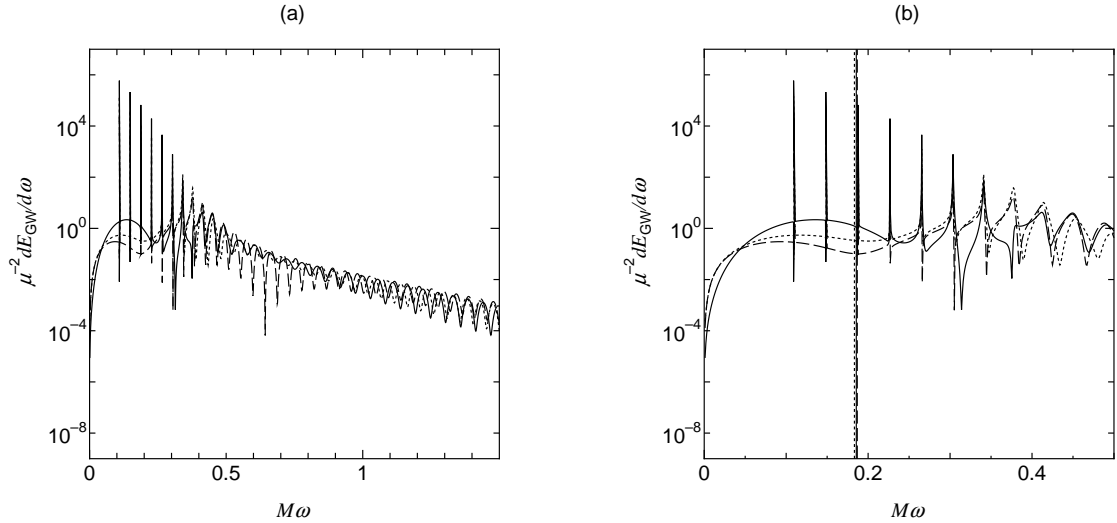}
\end{center}
\caption{
Energy spectrum ($l = 2$ mode) of gravitational waves from a spinning
particle with $E = 2.38 \mu$ and $L = 12.0 \mu M$ scattered by the
same star as Fig. \ref{fig:specfarrm2.26} ($R=2.26M$). In order to focus
on the detail of peaks in (a), we show the enlarged figure (b) for the
frequency range of  $0.0 \le M \omega \le 0.5$. Solid, dotted and
dashed lines show the case of the spin $S = 0$, $0.2\mu M$, and
$0.8\mu M$, respectively.
Vertical lines represent the angular velocity of a particle at the turning
point.
}
\label{fig:specnearrm2.26}
\end{figure}

\begin{figure}
\begin{center}
\leavevmode
\epsfysize=200pt
\epsfbox{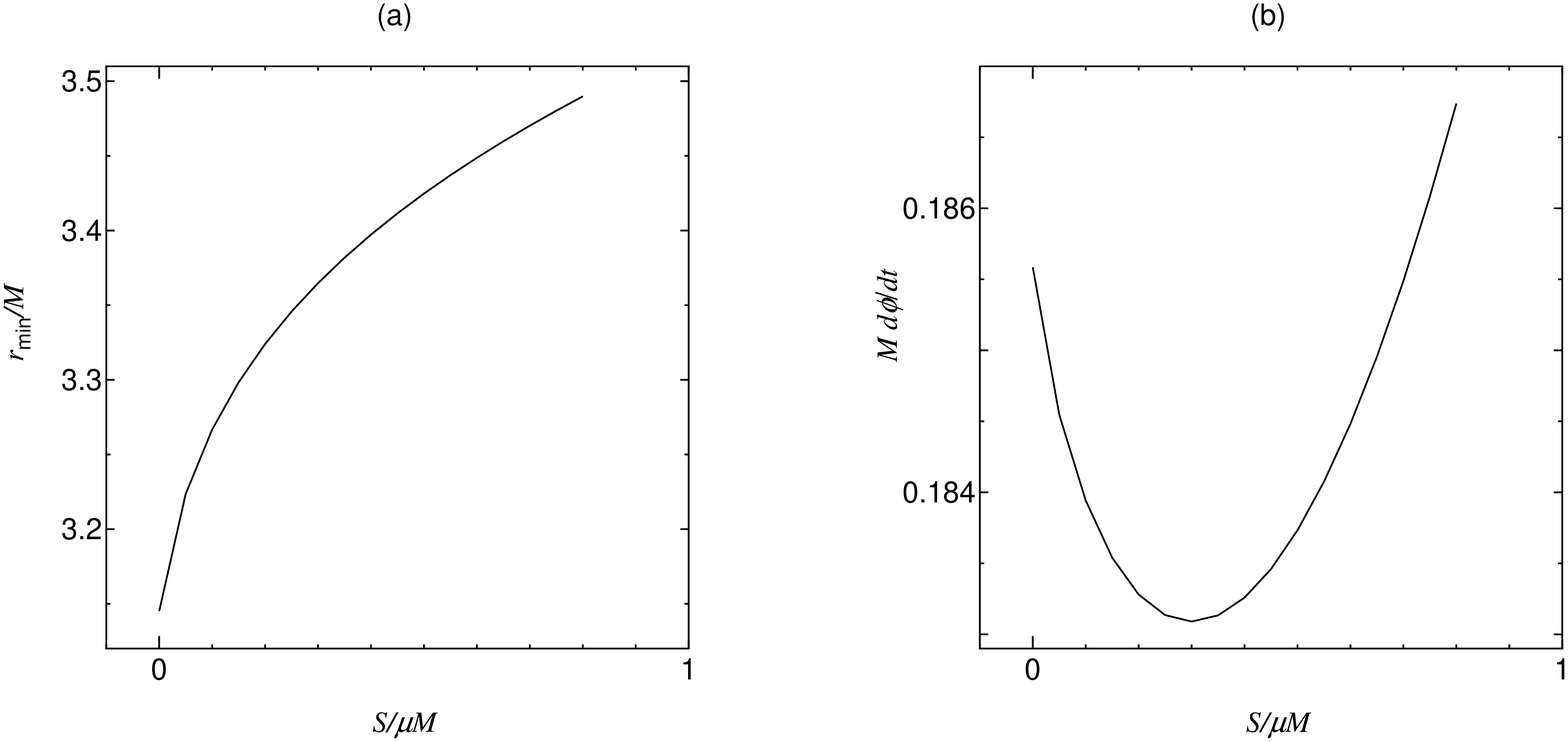}
\end{center}
\caption{
Location (a) and angular velocity (b) at the turning point
of a spinning particle in the case of 
$R = 2.26M$, $E = 2.38 \mu$, and $L = 12.0 \mu M$.
}
\label{fig:turnrm2.26_no2}
\end{figure}

If a test particle approaches very close to the stellar surface, e.g.
the turning point is $r_{\rm min} \simeq  3.15M \simeq 1.39R$ ($S = 0$)
and $r_{\rm min} \simeq 3.49M \simeq 1.54R$ ($S = 0.8 \mu M$) for a
particle with $E = 2.38 \mu$ and $L = 12.0 \mu M$, we find qualitative
difference in the spectrum (Fig. \ref{fig:specnearrm2.26}). 
Here, we note that in this case
a test particle with an  anti-parallel spin ($S < 0$) is not allowed to
give a scattered orbit, and any  particles with such parameters are
falling onto a star (More precisely, the lower limit of the spin value to
allow a  scattered orbit is $S \simeq -0.012 \mu M$). Thus we
discuss only the case of a parallel spinning particle. For a parallel spin
($S \ge 0$), the energy spectrum of gravitational waves is given in Fig.
\ref{fig:specnearrm2.26}.  As the spin increases, a broad peak in the
spectrum is monotonically shifted to the lower frequency region,
while the angular velocity at the turning point has the minimum point at some
spin $S \sim 0.3 \mu M$ (see Fig. \ref{fig:turnrm2.26_no2}).  To explain
this broad peak frequency, the turning point is no longer a good indicator,
although it is still one of the important factor to describe the emitted
gravitational waves.  Since the turning point is very close to
the surface of an ultracompact star, we suppose that there is another
factor relating to the effect of a strong gravitational field,
which probably affects to gravitational waves.  In fact, the location of a
turning point comes near to that of the maximum point of Regge-Wheeler
potential in this case.

\begin{figure}
\begin{center}
\leavevmode
\epsfysize=200pt
\epsfbox{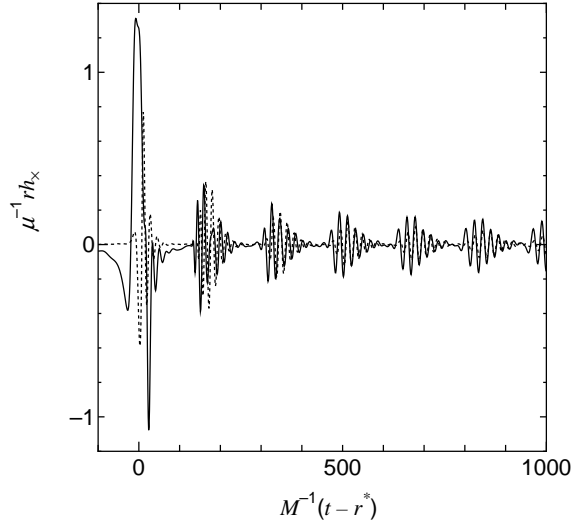}
\end{center}
\caption{
Waveform ($l = 2$ mode) of gravitational waves
in the direction of $\theta = \pi / 2$, $\phi = 0$
in the same situation as Fig. \ref{fig:specnearrm2.26}.
Solid and dotted lines show
the case of the spin $S = 0$ and $0.8\mu M$, respectively.
}
\label{fig:wfnearrm2.26}
\end{figure}

Since quasi-normal modes are excited by a test particle and trapped in
the potential well, those modes will leak periodically after the
encounter. In fact in Fig. \ref{fig:wfnearrm2.26}, we find  that
after the burst wave the ringing waves are emitted first at $t - r^{*} \sim 100M$, and those are repeated with a smaller amplitude than the
preceding one.  Since quasi-normal modes are independent 
of the
character of a particle, we do not find any dependence of the spin in
those ringing phase except for its amplitude.

\begin{figure}
\begin{center}
\leavevmode
\epsfysize=200pt
\epsfbox{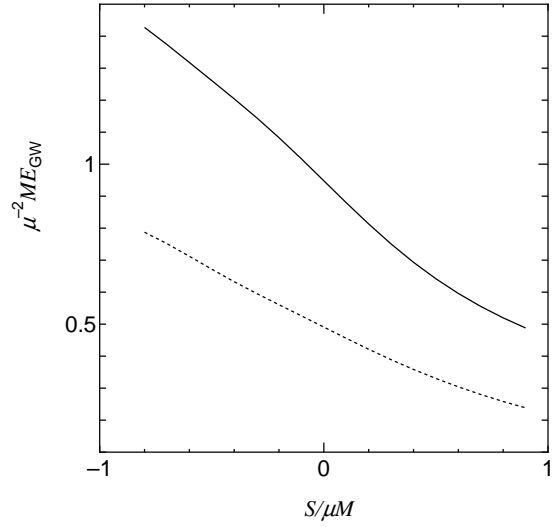}
\end{center}
\caption{
Total energy of gravitational waves
from a spinning particle with $E = 2.30 \mu$ and $L = 12.0 \mu M$
scattered by a uniform density star with $R = 2.26M$
for each multipole mode $l$.
Solid and dotted lines show
the case of $l = 2, 3$, respectively.
}
\label{fig:totalerm2.26}
\end{figure}

The total energy of the emitted gravitational waves are shown in 
Fig. \ref{fig:totalerm2.26} for a particle with $E = 2.30 \mu$ and $L =
12.0 \mu M$.  Here we show the total energy only for $l = 2, 3$,
because if the multipole mode $l$ becomes higher, some of the imaginary
part of quasi-normal modes become extremely small.  The existence
of these small imaginary parts requires to use very small step size of the
frequency in order to integrate the energy spectrum precisely, and
then the analysis becomes quite difficult and time consuming.  We also
believe that those two multipole modes are dominant and then the total energy
will not change so much (e.g. for the normal neutron star in Fig. \ref{fig:totalerm5.0},
the energy of $l = 4$ is about $10\%$ to the total sum energy).  We find that the
total energy similarly depends on the spin to the result for a normal
neutron star. The amount of the emitted energy increases as a spin is
large negative (anti-parallel), on the other hand  it will be smaller for
a large positive (parallel) spin. This result is similar to that for a normal neutron
star, but  the reason is not simple in this case. In the case of a
normal neutron star, we could easily
explain its spin dependence by considering a turning point.
 Because the location of the turning point monotonically
changes with the spin (Fig. \ref{fig:turnrm5.0}), then the gravitational
waves would also be  more radiated near the turning point as the spin
gets large negative (anti-parallel). However for the case of an ultracompact star,
from Fig. \ref{fig:turnrm2.26} we find that the location of the turning point has a
minimum. Hence the change of the turning point does not explain the
monotonic spin dependence of the total energy.  We should point out that since  the particle
passes through  very strong gravitational field, the coupling  between
an orbital angular momentum $\vec{L}$ and a spin
$\vec{S}$ is not simple, but may depend on the higher order terms of $M/r$
in Eq. (\ref{eqn:force}).

\begin{figure}
\begin{center}
\leavevmode
\epsfysize=200pt
\epsfbox{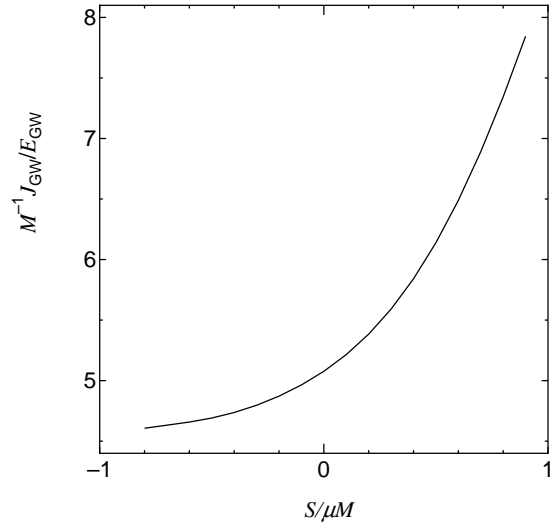}
\end{center}
\caption{
Ratio of the total angular momentum to the total energy radiated
$l = 2$ gravitational waves
in the same situation as Fig. \ref{fig:totalerm2.26}.
}
\label{fig:totaljerm2.26}
\end{figure}

For an ultracompact star, the ratio of the total angular momentum to the
total energy of  gravitational waves ($J_{\rm GW} / E_{\rm GW}$)
is shown in Fig. \ref{fig:totaljerm2.26}.  
As we mentioned before, 
the characteristic frequency of the system is important to think $J_{\rm GW} / E_{\rm GW}$.
However, for an ultracompact star 
many quasi-normal modes are excited by a scattered particle.
Therefore, we must  roughly estimate $(E_{\rm GW})_{l}$ and
$(J_{\rm GW})_{l}$ by using two frequencies:
One is the orbital frequency,
and the other is the quasi-normal mode frequencies.
Using these
two characteristic frequencies, $(E_{\rm GW})_{l}$ and $(J_{\rm
GW})_{l}$ can be described formally as 
\begin{eqnarray}
(E_{\rm GW})_{l}^{\rm (axial)} &\sim& 
(E^{(\rm cont)})_{l} + \sum_{i} (E^{(\rm qnm)}_{i})_{l},
\\ 
(J_{\rm GW})_{l}^{\rm (axial)} &\sim& 
(J^{(\rm cont)})_{l} + \sum_{i} (J^{(\rm qnm)}_{i})_{l}, 
\end{eqnarray}
where $(E^{(\rm cont)})_{l}$ and $(J^{(\rm cont)})_{l}$ are the total
energy and  angular momentum from the continuous spectrum, and
$(E^{(\rm qnm)}_{i})_{l}$ and $(J^{(\rm qnm)}_{i})_{l}$ are ones 
due to the $i$-th  quasi-normal mode.
We expect that  each contribution has the same relation
(\ref{eq:J_E_relation}),  i.e.,
\begin{eqnarray}
\left(\frac{J^{(\rm cont)}}{E^{(\rm cont)}}\right)_{l}^{\rm (axial)}
&\sim&
\frac{1}{\omega^{\rm (cont)}},
\\
\left(\frac{J^{(\rm qnm)}_{i}}{E^{(\rm qnm)}_{i}}\right)_{l}^{\rm (axial)} &\sim&
\frac{1}{\omega^{\rm(qnm)}_{i}}.
\label{eq:je_contrib_qnm}
\end{eqnarray}
Since $\omega^{\rm(qnm)}_{i}$ is independent of $S$, 
we believe that the spin dependence in Fig. \ref{fig:totaljerm2.26}
is only from the continuous spectrum, whose
characteristic frequency contains two parts:
One is the orbital effect which corresponds to the angular velocity at
the turning point and the other is the strong gravitational field effect 
which is excited to the motion of the particle.


\section{Concluding remarks}

\label{sec:conclusion}

We have studied axial modes of gravitational waves from a spinning
particle scattered by a uniform density star. Particularly, we have
focused on a  spin effect to the energy spectrum, the waveforms, the
total energy and angular momentum of gravitational waves, and
analyzed a dependence of the emitted gravitational waves on a particle
spin.

In the case of a normal neutron star model ($R \gtrsim 3M$) to which
an observable neutron star belongs,  we find that the total energy of $l
= 2$ gravitational waves gets  larger as the spin increases in the
anti-parallel direction to the orbital angular momentum.  For higher
multipole modes, however,  it is found that the total energy does not
depend monotonically  on the spin, but has a maximum.  The energy
spectrum have only one broad peak induced by the particle encounter.
This broad peak is affected by a spin of a particle, because the
trajectory is determined not only by the energy $E$ and the orbital
angular momentum $L$, but also by  the spin $S$. In fact, we find that
the broad peak of the energy spectrum shifts to the higher frequency
region, as the spin gets large negative (anti-parallel). This is because 
the angular velocity of a particle at the turning point increases
monotonically as the spin gets large negative (anti-parallel).  However,
it is known that this tendency of the broad peak shift is  also found
by a change of the orbital angular momentum. Therefore for a normal
neutron star model, it is very difficult to distinguish a spin effect from
the orbital angular momentum effect in the energy spectrum of axial
modes.

For an ultracompact star model ($R \lesssim 3M$), in addition to such 
an orbital contribution, we find sharp peaks in the energy
spectrum, which completely correspond  to the quasi-normal modes
excited by the scattered particle. If a particle passes through very close
to an ultracompact star ($r_{\rm min} \sim 3M$),  the trapped
quasi-normal modes give a prominent contribution to the total energy of
the emitted gravitational waves. For example, in an ultracompact star with
$R = 2.26M$, first eight trapped quasi-normal modes dominate.  In
particular in the case of Fig.
\ref{fig:totalerm2.26}, the contribution  from
fifth to eighth peaks gives about 45 $\sim $ 70 \% of the total energy. 
But if the turning point is far from a ultracompact star,  the
contribution from trapped quasi-normal modes is less than 10 \% 
(Fig. \ref{fig:totalerm5.0}). Note that in the case of
a particle passing through very close to an ultracompact star,
the frequency of a broad peak shifts monotonically as a spin increases, 
however the turning point and its
angular frequency do not  change monotonically  and have  the minimum
value at some value of the spin.
Therefore the interpretation of the cause of a broad
peak is not simple. 
This unusual dependence of the turning point on the spin
might be caused by a complicated coupling  between an orbital angular
momentum
$\vec{L}$ and a spin $\vec{S}$, 
and this coupling may appear when  a particle passes through
in the strong gravitational field near the maximum of Regge-Wheeler 
potential.

With respect to the ratio of the total angular momentum $J_{\rm GW}$ to 
the total energy $E_{\rm GW}$ of the emitted gravitational waves,
we could understand that its dependence on the spin is determined  by
the characteristic frequency, i.e., the angular velocity at the turning
point or the frequency of quasi-normal modes.
If the particle passes through the strong gravitational field, 
the ratio is also depend on 
the excitation of the strong gravitational field.
Since the quasi-normal modes do not depend on
the properties of a particle,  the dependence of the ratio $J_{\rm
GW}/E_{\rm GW}$ on the spin is due to the change of the trajectory.

From our results, we may conclude that the spin effects on the emitted
gravitational waves are seen only as the orbital change of a
spinning particle due to a spin-orbit interaction in
Eq. (\ref{eq:spin_eom_p}) and Eq. (\ref{eq:spin_eom_s}).
However, we expect another spin effect, which is a direct change of
the source term of gravitational waves via a spin contribution on the
energy-momentum tensor (\ref{eq:spinparticle_emtensor}).  This may
affect the emitted gravitational waves and its change could extract a
characteristic property of a spin from the observed gravitational
waves.  However, in the case of a spinning test particle, our results
show that the spin effect is mostly found as the orbital change by the
equations of motion. In fact, we expand the source term
with the value of the spin $s$ for the first order in order to examine the
particle spin contribution to the spectrum from the energy-momentum
tensor.  We compute the spectrum in the case of the source term with
no spin, with first order spin, and with full order spin.  There are
no remarkable difference among these three results, and all of them has a
global peak near the turning point for the normal neutron star.
Therefore, we conclude that the spin effect via the spin part of  the
energy-momentum tensor  is much  less than that by the orbital change.
In a realistic binary neutron star, however, a spin part of the
energy-momentum tensor plays definitely an important role in the
gravitational wave, because it has a stellar structure and it is essentially different
from that of the point particle, and the emitted gravitational waves
may depend on such a term as well as the orbital change.

In a realistic situation,  the polar modes are also important,
because those contain the fluid oscillation modes and then  $f$, $p$
modes have some information about high density fluid matter of a
neutron star.  Then the effect of a spin on those modes may reveal
more information about neutron star matter.  Although we could analyze
those modes in the present model, that is a perturbation approach with
a spinning test particle, we expect that the results will be negative.
Because the different interaction between a particle spin and polar
perturbations from axial perturbation will appear mainly in the
energy-momentum tensor.
However, as mentioned above, since the contribution from the
energy-momentum tensor would be very small in the perturbation
approach, we believe that the result for the polar modes may be
similar to the axial case,  i.e., the main factor of the spin effect
is caused by the orbital change and the difference of $f$, $p$ modes
due to the particle spin appears only in an enhancement of its
strength.  On the other hand, if a star is rotating, we expect a
coupling between a stellar rotation and a spin of a particle,
and then it is important to  analyze it, particularly because
a rotating star shows the $r$ mode instability.

\acknowledgments

MS would like to thank  Center for Gravitational Physics and
Geometry, the Pennsylvania State University and Department of
Physics, University of Illinois at Urbana-Champaign for their
hospitality, where part of this work was done.   KM is grateful to the
Albert-Einstein-Institut (Potsdam) for their hospitality, where part of
this work was done.  The numerical computation was mainly performed
by the NEC-SX4 vector computer at Yukawa Institute for Theoretical
Physics and the FUJITSU-VX vector computer at Media Network Center,
Waseda University.  This work was supported partially by a JSPS
Grant-in-Aid (Nos. 095689 and 1205705, and Specially Promoted
Research No. 08102010), and by the Waseda University Grant for
Special Research Projects.

\appendix

\section{Source term of a spinning particle for the axial mode}
\label{sec:sourceterm}

For a spinning particle, the difference in the analysis from our
previous work (non-spinning particle) \cite{TominagaSaijoMaeda}
appears only in the source term ${\cal S}_{lm\omega}^{\rm (ext)}$ in
Eq. (\ref{eq_ext}).  Since the detail of our method is shown in Ref
\cite{TominagaSaijoMaeda}, we only present the part of the
calculation concerned with the source term of a spinning particle.

The amplitude of the gravitational wave is given by Eq. (A24) in Ref
\cite{TominagaSaijoMaeda} and then we must calculate the following
integration,
\begin{eqnarray}
\int_{\hat{r}_{\rm min}^{*}}^{\infty} dr^{*}
X^{({\rm ext})(0)}_{lm\omega}(r^{*}) 
{\cal S}_{lm\omega}^{\rm (ext)}(r^{*}).
\label{eqn:Sintegral}
\end{eqnarray}
When a spinning particle is scattered on the equatorial plane by a 
relativistic star or a black hole, with its trajectory as
$z^\mu(\tau)=(\hat{t}(\tau),
\hat{r}(\tau),
\hat{\theta}=\pi/2,
\hat{\phi}(\tau))$, 
we find  Eq. (\ref{eqn:Sintegral}) as follows:
\begin{eqnarray}
&& \int_{\hat{r}_{\rm min}^{*}}^{\infty} dr^{*}
X^{({\rm ext})(0)}_{lm\omega}(r^{*}) 
{\cal S}_{lm\omega}^{\rm (ext)}(r^{*}) 
\nonumber
\\ && \qquad =  \frac{4 \pi\mu}{n + 1} \left[ -2 i C_{lm} \left. 
\frac{d P_{lm}}{d\theta} 
\right|_{\theta = \pi / 2}\right] \int_{0}^{\infty} d \hat{t} 
K_{S} \left( \frac{d \hat{t}}{d \tau} \right)^{-1} 
\nonumber \\
&& \qquad \qquad 
\times \left(\hat{Q}^{(0)}_{m\omega} X^{({\rm ext})(0)}_{lm\omega} 
\left( \hat{t} \right)  + \hat{Q}^{(1)}_{m\omega} 
\frac{dX^{({\rm ext})(0)}_{lm\omega}}{d \hat{r}}
\left( \hat{t} \right) \right) \nonumber \\
&& \qquad \qquad + \frac{4 \pi\mu }{n(n + 1)} 
\left[2 i m C_{lm} \left.
\frac{d P_{lm}}{d \theta} \right|_{\theta = \pi / 2}
\right]
\nonumber \\
&&
\qquad \qquad \times \int_{0}^{\infty} d \hat{t}  K_{S} 
\left( \frac{d \hat{t}}{d \tau} \right)^{-1} 
\left(\hat{D}^{(0)}_{m\omega} X^{({\rm ext})(0)}_{lm\omega} 
\left( \hat{t} \right)
\right.
\nonumber \\
&&
\qquad \qquad 
\left.
 + \hat{D}^{(1)}_{m\omega} \frac{dX^{({\rm
ext})(0)}_{lm\omega}}{d \hat{r}}
\left( \hat{t} \right)
+ \hat{D}^{(2)}_{m\omega} \frac{d^{2} X^{({\rm ext})(0)}_{lm\omega}}{d \hat{r}^{2}} 
\left( \hat{t} \right)
\right).
\end{eqnarray}
where
$n = l(l + 1)/2 - 1$, and $C_{lm}$ is a normalization constant 
of a spherical harmonics $Y_{lm}$, written as
\begin{equation}
C_{lm} = (-1)^{(m + |m|)/2} 
\sqrt{\frac{(2l + 1)}{4 \pi} \frac{(l - |m|)!}{(l + |m|)!}}.
\end{equation}
$P_{lm}$ is the associated Legendre function, and 
$\hat{Q}^{(0)}_{m\omega}$, $\hat{Q}^{(1)}_{m\omega}$, 
$\hat{D}^{(0)}_{m\omega}$, $\hat{D}^{(1)}_{m\omega}$ and $\hat{D}^{(2)}_{m\omega}$
are given as
\begin{eqnarray}
\hat{Q}^{(0)}_{m\omega} &=& \hat{r}^2 
\left( \frac{1}{\hat{r}^3 - M \tilde{S}^2} + 
\frac{1}{\hat{r}^3 + 2 M \tilde{S}^2}
\right)
\frac{d \hat{r}}{d \tau} \frac{d \hat{\phi}}{d \tau} 
\nonumber \\
&& \times
\sin \left( \omega \hat{t} - m  \hat{\phi} \right) 
\nonumber \\ 
&& + \tilde{S} \Biggl[ 
\omega \Biggl\{ \frac{\hat{r}^2}{(\hat{r} - 2M) 
(\hat{r}^3 - M \tilde{S}^2)} 
\left( \frac{d \hat{r}}{d \tau} \right)^2  
\nonumber \\
&& 
- \frac{\hat{r}^3}{\hat{r}^3 + 2 M \tilde{S}^2} 
\left( \frac{d \hat{\phi}}{d \tau} \right)^2 
\Biggr\} \cos \left( \omega \hat{t} - m \hat{\phi} \right) 
\nonumber \\
&& 
+ m \frac{\hat{r} - 2M}{\hat{r}^3 - M \tilde{S}^2}  \frac{d
\hat{t}}{d \tau} \frac{d \hat{\phi}}{d
\tau}
\cos \left( \omega \hat{t} - m \hat{\phi} \right)
\Biggr]
, \\
\hat{Q}^{(1)}_{m\omega} &=& 
\tilde{S} \frac{\hat{r} - 2M}{\hat{r}^3 -  M \tilde{S}^2} 
\frac{d \hat{t}}{d \tau} \frac{d \hat{r}}{d \tau}
\sin \left( \omega \hat{t} - m \hat{\phi} \right)
, \\
\hat{D}^{(0)}_{m\omega} &=& 
\frac{\hat{r}^2 (\hat{r} - 2M)}{\hat{r}^3 + 2Ms^2} 
\left( \frac{d \hat{\phi}}{d \tau} \right)^2
\cos \left( \omega \hat{t} - m \hat{\phi} \right)
\nonumber \\
&&
+ \tilde{S} \Biggl[
- \omega \frac{\hat{r}^2}{\hat{r}^3 - M \tilde{S}^2} \frac{d \hat{r}}{d \tau} 
\frac{d \hat{\phi}}{d
\tau}
\sin \left( \omega \hat{t} - m \hat{\phi} \right) 
\nonumber \\
&& 
- \frac{(\hat{r} - 4 M) (\hat{r} - 2M)}{\hat{r} (\hat{r}^3 -  M
\tilde{S}^2)} \frac{d \hat{t}}{d \tau} \frac{d \hat{\phi}}{d
\tau}
\cos \left( \omega \hat{t} - m \hat{\phi} \right)
\Biggr]
, \\
\hat{D}^{(1)}_{m\omega} &=& 
\frac{\hat{r}^3 (\hat{r} - 2M)}{\hat{r}^3 + 2Ms^2} 
\left( \frac{d \hat{\phi}}{d \tau} \right)^2
\cos \left( \omega \hat{t} - m \hat{\phi} \right)
\nonumber \\
&&
+ \tilde{S} \Biggl[
- \omega \frac{\hat{r}^3}{\hat{r}^3 - M \tilde{S}^2} 
\frac{d \hat{r}}{d \tau} \frac{d \hat{\phi}}{d \tau}
\sin \left( \omega \hat{t} - m \hat{\phi} \right) 
\nonumber \\
&&
+ \frac{\hat{r} (\hat{r} - 2M)}{\hat{r}^3 - M \tilde{S}^2}  
\frac{d \hat{t}}{d \tau} \frac{d \hat{\phi}}{d \tau}
\cos \left( \omega \hat{t} - m \hat{\phi} \right)
\Biggr]
, \\
\hat{D}^{(2)}_{m\omega} &=&
\tilde{S} \frac{\hat{r} (\hat{r} - 2M)^2}{\hat{r}^3  - M \tilde{S}^2}
\frac{d \hat{t}}{d \tau} \frac{d \hat{\phi}}{d \tau}
\cos \left( \omega \hat{t} - m \hat{\phi} \right). 
\end{eqnarray}
$K_{S}$ is written as
\begin{eqnarray}
K_{S} = 1 - \frac{M \tilde{S}^2}{r^3} \left[ 1 + 3 \left( u^{\phi} r
\sin \theta \right)^2 \right],
\end{eqnarray}
which is defined in
the relation between the 4-velocity $v^{\mu}$ and the
normalized momentum $u^{\mu}$ as
\begin{eqnarray}
K_{S} v^{t} &=& \left( 1 - \frac{M \tilde{S}^2}{r^3} \right) u^{t}, \\
K_{S} v^{r} &=& \left( 1 - \frac{M \tilde{S}^2}{r^3} \right) u^{r}, \\
K_{S} v^{\phi} &=& \left( 1 + \frac{2 M \tilde{S}^2}{r^3} \right)
u^{\phi}.
\end{eqnarray}


\newpage

\twocolumn[\hsize\textwidth\columnwidth\hsize\csname
@twocolumnfalse\endcsname

\vskip2pc]

\twocolumn[\hsize\textwidth\columnwidth\hsize\csname
@twocolumnfalse\endcsname

\vskip2pc]

\twocolumn[\hsize\textwidth\columnwidth\hsize\csname
@twocolumnfalse\endcsname

\vskip2pc]

\twocolumn[\hsize\textwidth\columnwidth\hsize\csname
@twocolumnfalse\endcsname

\vskip2pc]

\twocolumn[\hsize\textwidth\columnwidth\hsize\csname
@twocolumnfalse\endcsname

\vskip2pc]

\twocolumn[\hsize\textwidth\columnwidth\hsize\csname
@twocolumnfalse\endcsname

\vskip2pc]

\twocolumn[\hsize\textwidth\columnwidth\hsize\csname
@twocolumnfalse\endcsname
\vskip2pc]


\begin{thebibliography}{00}
\bibitem{footnote1} We use a tilde for variables normalized by a
particle mass $\mu$.
\end{thebibliography}
\end{document}